\nonstopmode
\documentclass[apl,twocolumn,reprint,nofootinbib]{revtex4-1}
\usepackage{graphicx}
\usepackage[font=small,justification=raggedright,labelfont=bf]{caption}

\usepackage{xcolor}
\usepackage{xspace}
\usepackage{comment}

\usepackage{amsmath, amssymb}
\usepackage{subfig}

\newcommand{\figureref}[1]{Fig.~\ref{#1}}
\newcommand{\Iarray}{I_{\mathrm{array}}\xspace}

\begin{document}
	
\title{Computer-automated tuning of semiconductor double quantum dots into the single-electron regime}
\author{T.~A. Baart$^{1,2}$}
\author{P.~T. Eendebak$^{1,3}$}
\author{C. Reichl$^{4}$}
\author{W. Wegscheider$^{4}$}
\author{L.~M.~K. Vandersypen$^{1,2}$}
\affiliation{1: QuTech, Delft University of Technology, P.O. Box 5046, 2600 GA Delft, The Netherlands}
\affiliation{2: Kavli Institute of Nanoscience, Delft University of Technology, P.O. Box 5046, 2600 GA Delft, The Netherlands}
\affiliation{3: Netherlands Organisation for Applied Scientific Research (TNO), P.O. Box 155, 2600 AD Delft, The Netherlands}
\affiliation{4: Solid State Physics Laboratory, ETH Z\"{u}rich, 8093 Z\"{u}rich, Switzerland}
\date{\today}

\begin{abstract}
	We report the computer-automated tuning of gate-defined semiconductor double quantum dots in GaAs heterostructures. We benchmark the algorithm by creating three double quantum dots inside a linear array of four quantum dots. The algorithm sets the correct gate voltages for all the gates to tune the double quantum dots into the single-electron regime. The algorithm only requires (1) prior knowledge of the gate design and (2) the pinch-off value of the single gate $T$ that is shared by all the quantum dots. This work significantly alleviates the user effort required to tune multiple quantum dot devices.
\end{abstract}

\maketitle
	
\begin{figure}[tb]
	\includegraphics[width=0.5\textwidth]{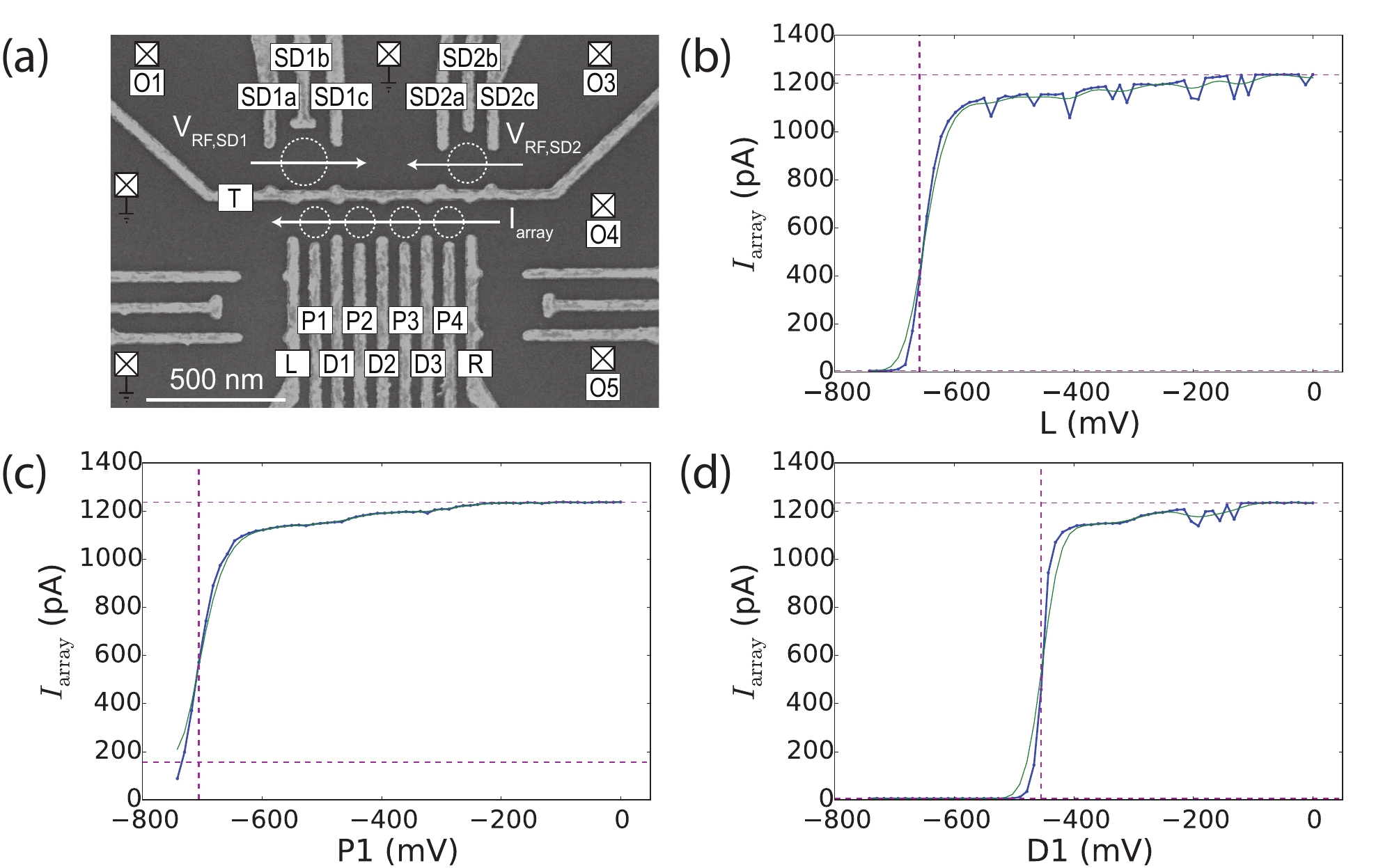} 
	\caption{(a) SEM image of a sample nominally identical to the one used for the measurements. Dotted circles indicate quantum dots, squares indicate Fermi reservoirs in the 2DEG, which are connected to ohmic contacts. O5 is always open during the measurements. The gates that are not labeled are grounded. The current through the quadruple dot array, $\Iarray$, is measured to create single dots. The reflectance of the SDs, $V_{RF,SD1}$ and $V_{RF,SD2}$, is monitored to tune DQDs into the single-electron regime. (b-d) Pinch-off curves showing $\Iarray$ versus the gate voltages $L$, $P1$ and $D1$ ($T=-400$~mV). The calculated transition value is shown as a dotted vertical line. The horizontal lines denote the high and low value as detected by the algorithm (see Supplementary Information~\ref{sec:pinch_off}).}
	\label{fig:Fig1}
\end{figure}

Electrostatically defined semiconductor quantum dots have been the focus of intense research for the application of solid-state quantum computing~\cite{Hanson2007,Zwanenburg2013,Kloeffel2013}. In this architecture, quantum bits (qubits) can be defined by the spin state of an electron. Recently, several experiments have shown coherent manipulation of such spins for the purpose of spin-based quantum computation~\cite{Petta2005,Nowack2007,Medford2013,Kawakami2014,Veldhorst2014}. Enabled by advances in device technology, the number of quantum dots that can be accessed is quickly increasing from very few to many \cite{Thalineau2012,Takakura2014}. Up to date, all these quantum dots have been tuned by `hand'. This is a slow process whereby gate voltages are tweaked carefully, first to reach a regime with one electron in each of the dots, and then to adjust the strength of all the tunnel barriers. Defects and variations in the local composition of the heterostructure lead to a disordered background potential landscape, which must be compensated for by the gate voltages. On top, cross-capacitances of each gate to neighboring dots increases the tuning complexity as the number of dots increases. The ability to tune these dots automated by computer algorithms, including tuning of many dots in parallel, is an important ingredient towards the scalability of this approach to create a large-scale quantum computer.\\
\indent In this Letter, we demonstrate the computer automated tuning of double quantum dot (DQD) devices. We have created an algorithm that only requires as input: (1) prior knowledge of the gate design, which is reasonable for future large-scale quantum dot circuits and (2) the measured pinch-off value of the single gate $T$ shared by all the quantum dots. We describe the algorithm used and verify its robustness by creating three independent DQDs inside a quadruple dot array. The algorithm finds the correct gate voltages to tune all DQDs into the single-electron regime and the computer recognizes that this goal has been achieved within an overnight measurement.\\
\indent A scanning electron microscopy (SEM) image of a device nominally identical to the one used is shown in Fig.~\ref{fig:Fig1}(a). Gate electrodes fabricated on the surface of a GaAs/AlGaAs heterostructure are biased with appropriate voltages to selectively deplete regions of the two-dimensional electron gas (2DEG) 90~nm below the surface and define the quantum dots. The main function of each gate is as follows: gates $L$ and $R$ set the tunnel coupling with the left and right reservoir, respectively. $D1-D3$ control the three inter-dot tunnel couplings and $P1-P4$ are used to set the electron number in each dot. However, each gate influences the other parameters as well. Changing $L$ for example, will also change the electron number in dot 1 and influence the inter-dot tunnel barrier between dot 1 and 2. This needs to be taken into account by the algorithm. Two other nearby quantum dots on top of the qubit array, sensing dot 1 and 2 (SD1 and SD2), are created in a similar way and function as a capacitively coupled charge sensor of the dot array. When positioned on the flank of a Coulomb peak, the conductance through the sensing dot is very sensitive to the number of charges in each of the dots in the array. Changes in conductance are measured using radiofrequency (RF) reflectometry \cite{Barthel2010}.
High-frequency lines are connected via bias-tees to gates $P1$, $P3$ and $P4$. The device was cooled inside a dilution refrigerator to a base temperature of $\sim$15 mK. All measurements were taken at zero magnetic field. \\
\indent Before running the algorithm the user is required to input a range of $T$-values for which the algorithm should try to find DQDs. This range is currently determined by measuring the pinch-off value of $T$ manually, and then choosing a set of gate voltages more negative than this pinch-off value. The pinch-off value can for example be determined by setting all other gates to 0~mV and next measuring the current from O1 to O4 (other ohmics open) whilst sweeping $T$. This step could be automated in future work.\\ 
\indent The algorithm consists of 3 steps: (1) to determine the starting values for the gate voltages, we first measure the pinch-off characteristic between each individual gate and the shared $T$-gate. Based on those results we (2) create single quantum dots. The required tunnel barriers acquired in (2) can be used as a starting point to (3) create double dots into the single-electron regime. Subsequently, steps (1) and (2) are used to create the SDs.\\
\indent To measure the pinch-off characteristic we apply a small voltage-bias ($\sim500~\mu$V) to O4 and measure the current $\Iarray$ through the quadruple dot array. Variations in the local composition of the heterostructure underneath each gate will be reflected in the required voltage to create quantum point contacts (QPCs). We term this voltage the transition value, $V_{\mathrm{gate},i}^{\mathrm{tr}}$, which is defined as the gate voltage for which $\Iarray$ is at $\sim30$\% of its maximum value (see Supplementary Information~\ref{sec:pinch_off}). This procedure is repeated for a range of $T$-values. Figs.~\ref{fig:Fig1}(b-d) show an example for {$T=-400$~mV} and the gates controlling the leftmost dot ($L$, $P1$ and $D1$). In practice, it is best to continue with the most positive $T$-value that still allows pinch-off for all gates. In our experience this tends to create better quantum dots for this gate design.\\ 
\indent We start by creating single quantum dots, as they already include much of the cross-talk between gates, dots and barriers, that is present in double dots.
To create single quantum dots we apply a fixed voltage for the plunger gate (usually -80 mV) which we know is appropriate for this device design, and use the transition values of the barrier gates as input for a 2D coarse scan. A suitable scan range is [$V_{\mathrm{gate},i}^{\mathrm{tr}}$-10~mV, $V_{\mathrm{gate},i}^{\mathrm{tr}}$+400~mV]. We again monitor $\Iarray$. The structure of these scans is always similar: for negative gate voltages the channel is closed, so there is no current. For more positive voltages the channel is open, so there is a large current. We fit a tetragon to the area corresponding to large current, see Fig. ~\ref{fig:refinement}(a) for an example of the leftmost dot (details can be found in Supplementary Information~\ref{sec:analysis_of_single_dots}). We next take a finer scan of the area closest to the tetragon corner with the most negative gate voltages, see Figs.~\ref{fig:refinement}(b-e). In the experiments we have performed, this point is always showing the start of quantum dot formation through the appearance of a Coulomb peak. We use this point as the starting point in gate-space for creating DQDs. The exact location of the Coulomb peak is determined using a Gabor filter and is shown as black dots in Figs.~\ref{fig:refinement}(b-e) (see Supplementary Information~\ref{sec:analysis_of_single_dots}).\\ 
\indent When going to double dots, transport measurements are not suitable as current levels through few-electron double dots are impractically low for this device design. Therefore, once the single dots have been formed, we tune the SDs in a similar way. They can then be used for non-invasive charge sensing which does allow one to distinguish single-electron transitions in the dot array through RF-reflectometry. To achieve a high sensitivity it is important that the SD is tuned to the flank of one of its Coulomb peaks. After finding a Coulomb peak for the SD in a similar way as described for the qubit dots, we make a 1D scan of the plunger gates, see Fig.~\ref{fig:refinement}(f). Each detected Coulomb peak is given a score based on its height and slope that allows the algorithm to continue with the most sensitive operating point for the corresponding plunger gate (see Supplementary Information~\ref{section:coulumbtuning}).\\
\indent With the SD tuned we create a double dot in the following way: first we set the voltages of the gates for the double dot to the values found for the individual single dots (black dots in Figs.~\ref{fig:refinement}(b-e)). For the single gate shared by the two individual dots (e.g.~gate $D1$ for the leftmost double dot) the average of the two values is used. Next, we record a charge stability diagram of the double dot structure by varying the two plunger gate voltages involved. We use a heuristic formula to determine the correct scan range that takes into account the capacitive coupling of the gates to the dots (see Supplementary Information~\ref{sec:tuning_double_dot}). Typical results for such scans are shown in Fig.~\ref{fig:Fig3}(a-c). Scans involving two plungers are measured by applying a triangular voltage ramp to the plungers on the horizontal axis using an arbitrary waveform generator, and by stepping the other plunger gate using DACs~\cite{Baart2015}. Whilst stepping the latter gate we also adjust the sensing dot plunger gate to compensate for cross-capacitive coupling and thereby improve the operating range of the SD.\\
\indent To verify that the double dot has reached the single-electron regime, the algorithm first detects how well specific parts of the charge stability diagrams match the shape of a reference cross (see inset of Fig.~\ref{fig:Fig3}). Each match should ideally correspond to the crossing of a charging line from each dot. The shape of the reference cross is derived from the various capacitive couplings, which follow from the gate design and are known approximately from the start. Instead of detecting crosses, one could also try to detect the individual charge-transition lines. This turned out to be more sensitive to errors for two reasons: (1) Extra features in the charge stability diagrams that do not correspond to charging lines are wrongfully interpreted as dot features. (2) Not all charging lines are straight across the entire dataset; this makes it harder to interpret which line belongs to which dot. The cross-matching algorithm is robust against such anomalies because of the local, instead of global, search across the dataset. In future work it could actually be useful to still detect these extra and/or curved lines. They could give information about e.g. unwanted additional dots and aid in determining the electron numbers in regions with higher tunnel couplings. For the current goal of finding the single-electron regime this extra information is not required.\\
\indent Next, the algorithm checks whether within a region slightly larger than 70$\times$70~mV$^{2}$, it finds other charge transitions for more negative gate voltages with respect to the most bottom-left detected cross (see Supplementary Information~\ref{sec:tuning_double_dot}). These regions are depicted by the green tetragons in Fig.~\ref{fig:Fig3}. If no extra transitions are detected: the single-electron regime has been found and the result is given a score of 1 for that specific measurement outcome. If extra transitions are found the algorithm outputs the score 0. In both cases this is where the algorithm stops. At the end of the run the user can see the measurement results for the various initial choices of $T$ and select the best one.\\
\indent All combined, the running of this complete algorithm (for a single value of the $T$-gate) takes $\sim 200$~minutes. Per device typically 5 $T$-values are tested. In practice we have observed that for some cooldowns of the sample the algorithm could not attain the single-electron regime. A thermal cycle combined with different bias cooling~\cite{Long2006} can significantly influence the tuning and solve this issue; just as for tuning done by hand. The key difference is that with the computer-aided tuning hardly any user effort is required to explore tuning of double dots to the few-electron regime. In future work the time required for automated tuning (as well as for tuning by hand) can be further reduced by also connecting the tunnel barrier gates of each single dot to a high-frequency line which would allow much faster scans for Figs.~\ref{fig:Fig1}-\ref{fig:refinement}~\cite{Stehlik2015}. These scans currently form the bottleneck in the overall tuning process. Future experiments will also address the automated tuning of more than two dots and the tuning of the tunnel couplings in between dots and their reservoirs, which are key parameters for operating dots as qubit devices. \\
\indent In summary, we have demonstrated computer-automated tuning of double quantum dot devices into the single-electron regime. This work will simplify tuning dots in the future and forms the first step towards automated tuning of large arrays of quantum dots. 

\begin{acknowledgments}
The authors acknowledge useful discussions with the members of the Delft spin qubit team, and experimental assistance from M.~Ammerlaan, J.~Haanstra, R.~Roeleveld, R.~Schouten, M.~Tiggelman and R.~Vermeulen. This work is supported by the Netherlands Organization of Scientific Research (NWO) Graduate Program, the Intelligence Advanced Research Projects Activity (IARPA) Multi-Qubit Coherent Operations (MQCO) Program and the Swiss National Science Foundation.		
\end{acknowledgments}

\begin{figure*}[!htb]
	\centering
	\includegraphics[width=0.85\textwidth]{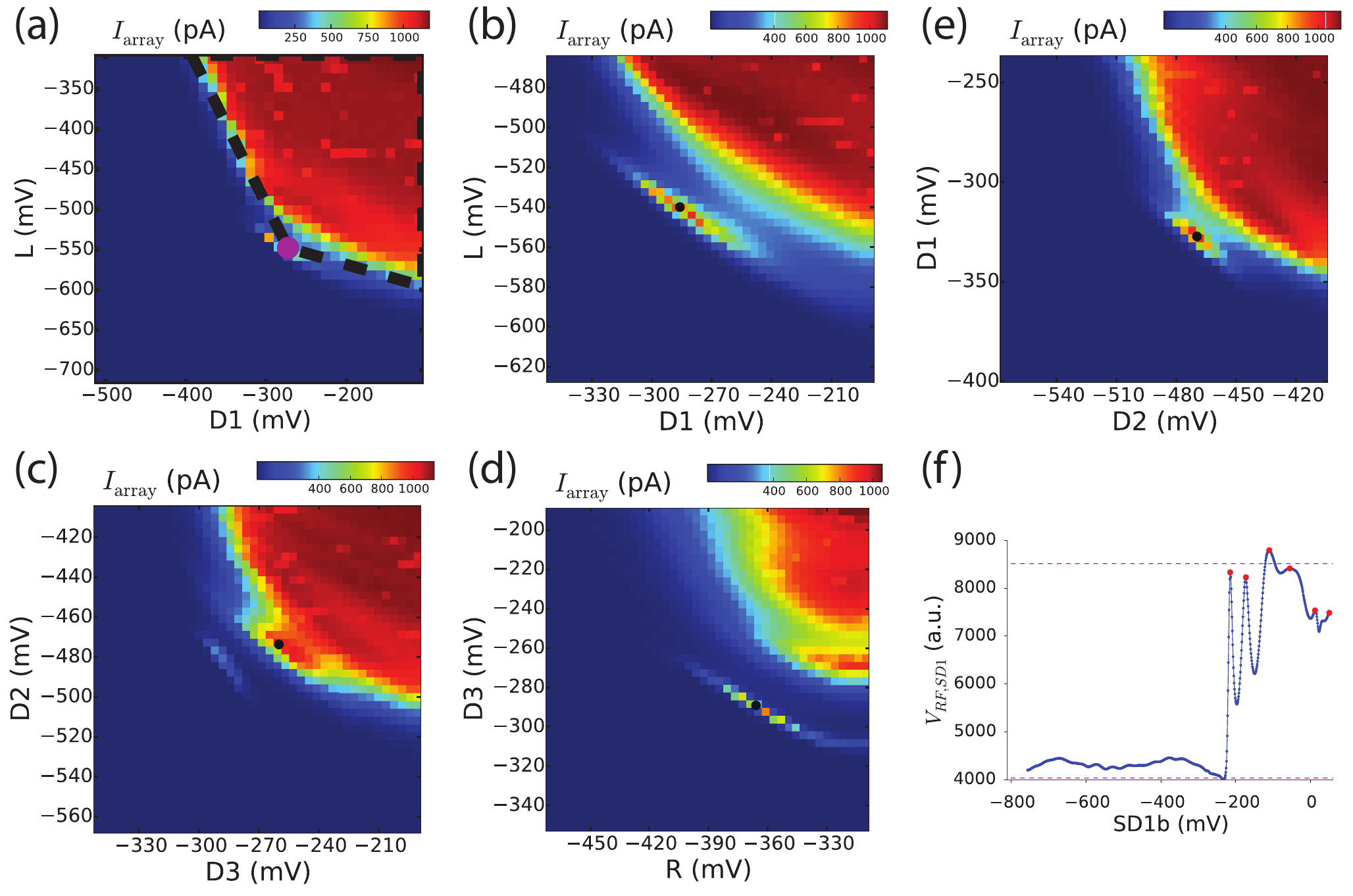}  
	\caption{(a-e) Four single dots are created by sweeping the tunnel barrier gate voltages for each dot whilst keeping the plunger gate fixed and monitoring $\Iarray$. After a coarse scan (example shown for the leftmost dot in (a)) we zoom in to the region showing Coulomb blockade (b-e). The center location where  Coulomb peaks are formed is determined using a Gabor filter and depicted by black dots. (f) The SD is fine-tuned by sweeping its plunger gate voltage. The charge sensing measurements shown in Fig.~\ref{fig:Fig3} are performed by tuning to the left flank of a Coulomb peak.}
	\label{fig:refinement}
\end{figure*}

\begin{figure*}[tb]
	\centering
	\includegraphics[width=0.95\textwidth]{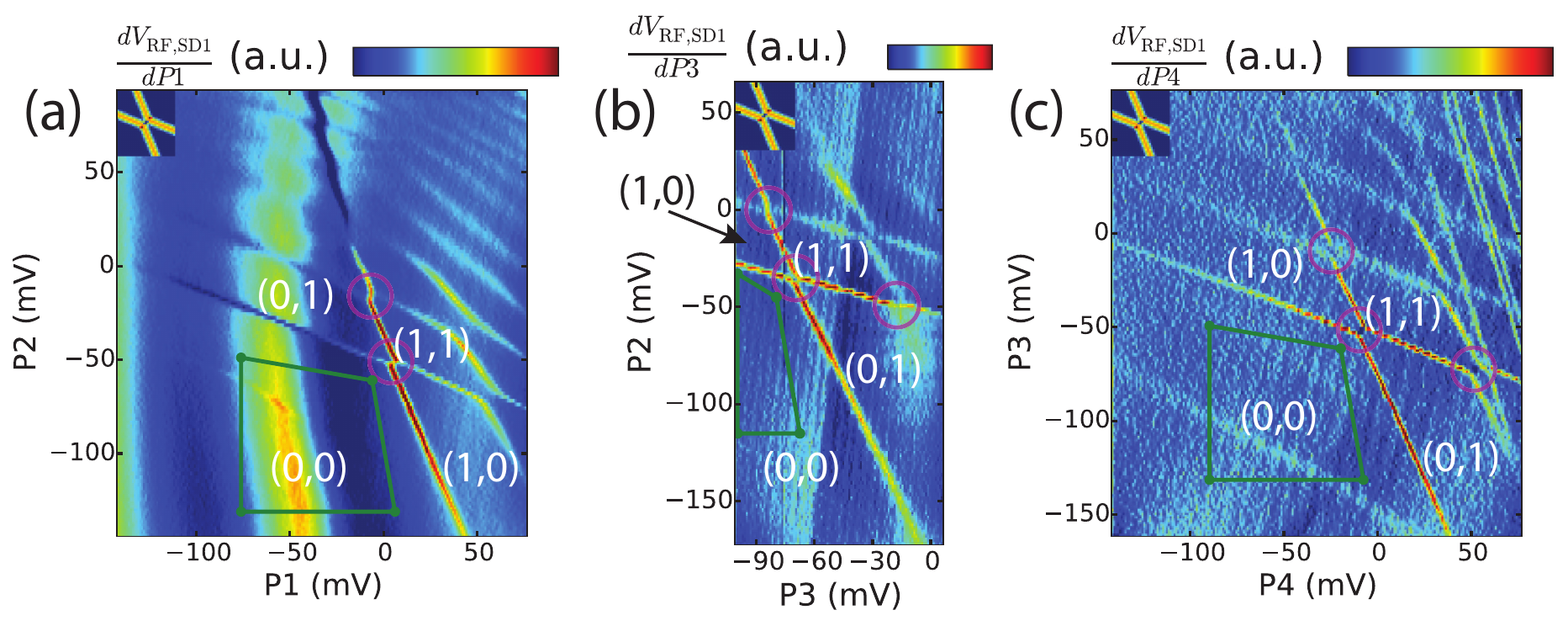} 
	\caption{ (a-c) Double dot charge stability diagram for the left, middle and right double dot respectively. Using the result of Figs.~\ref{fig:refinement}(b-e) the tunnel barriers are set, and the plunger gates are swept as indicated. The occupancy of each dot is denoted by $(n,m)$ corresponding to the number of electrons in the left and right dot respectively of that specific double dot. The algorithm determines how well regions of the charge stability diagram match to a reference cross (see inset). Good matches are encircled. These ideally corresponds to the crossing of charging lines from two dots. The single-electron regime is found by verifying that no other charging lines are observed for more negative gate voltages with respect to the most bottom-left detected cross (green regions). The horizontal scan range of panel (b) is less than for (a) and (c) due to hardware limitations.}
	\label{fig:Fig3}
\end{figure*}

\clearpage

\begin{center}
\textbf{Supplementary Material}
\end{center}

\section{Methods and materials}
The experiment was performed on a $\mathrm{GaAs/Al_{0.307}Ga_{0.693}As}$ heterostructure grown by molecular-beam epitaxy, with a 90-nm-deep 2DEG with an electron density of $\mathrm{2.2 \cdot 10^{11}\ cm^{-2}}$ and mobility of $\mathrm{3.4 \cdot 10^{6}\ cm^{2} V^{-1} s^{-1}}$ (measured at 1.3 K). The metallic (Ti-Au) surface gates were fabricated using electron-beam lithography. The device was cooled inside an Oxford Triton 400 dilution refrigerator to a  base temperature of 15 mK. To reduce charge noise the sample was cooled while applying a positive voltage on all gates (ranging between 100 and 400 mV) \cite{Long2006}. Gates $P1$, $P3$ and $P4$ were connected to homebuilt bias-tees ($RC$=470 ms), enabling application of d.c.~voltage bias as well as high-frequency voltage excitation to these gates. Frequency multiplexing combined with RF reflectometry of the SDs was performed using two LC circuits matching a carrier wave of frequency 107.1 MHz for SD1 and 86.4 MHz for SD2. The inductors are formed by microfabricated NbTiN superconducting spiral inductors with an inductance of 3.2 $\mu$H (SD1) and 4.6 $\mu$H (SD2). The power of the carrier wave arriving at the sample was estimated to be -93 dBm. The reflected signal was amplified using a cryogenic Weinreb CITLF2 amplifier and subsequently demodulated using homebuilt electronics. Data acquisition was performed using a FPGA (DE0-Nano Terasic) and digital multimeters (Keithley). Voltage pulses to the gates were applied using a Tektronix AWG5014.

\section{Software and algorithms}

The software was developed using Python~\cite{Python} with SciPy~\cite{SciPy}.

The image processing is performed in pixel coordinates. We specify the parameters of algorithms in physical units such as mV. The corresponding parameter in pixel units is then determined by translating the value using the scan parameters. By specifying the parameters in physical units the algorithms remain valid if we make scans with a different resolution. Of course making scans with a different resolution can lead to differences in rounding of numbers leading to slightly different results.

\subsection{Determination of the transition values}
\label{sec:pinch_off}
To determine the transition values we perform the following steps:
\begin{itemize}
	\item Determine the lowvalue ($L$) and highvalue ($H$) of the scan by taking a robust minimum and maximum. For $L$ this is done by taking the 1\textsuperscript{th} percentile of the values. $H$ is determined by first taking the 90\textsuperscript{th} percentile of the scan data $H_0$ and then the 90\textsuperscript{th} percentile of all the values larger then ($L+H_0$)/2. This two-stage process to determine $H$ also works well when the pinch-off occurs for very positive gate voltages. Simply taking for example the 99\textsuperscript{th} percentile of the scan data could then result in a too low estimate.	
	\item Smoothen the signal and find the first element in the scan larger than $.7 L + .3 H$. The position of that value is selected as the transition value.
	
	\item Perform several additional checks. The above two steps will always results in a transition value, even though the channel could be completely open or closed. 
	The checks include amongst others:
	\begin{itemize}
		\item If the transition value is near the left border of the scan we conclude the transition has not been reached. We then set the transition value to the lowest value of the gate that has been scanned. In principle the algorithm could continue to search for a transition value for more negative gate voltages. However, making gate voltages too negative may induce charge noise in the sample so we do not want to apply very negative voltages. Choosing the most negative voltage of the scan range then turns out to be a good choice. In the next steps of the algorithm, this transition voltage is  just a starting value and the gate voltage will still be varied. Due to cross-talk, the neighboring gates in follow-up steps will together with the gate that did not yet close, typically still ensure the formation of single dots.
	
		\item The difference of the mean of the measured values left of the transition value and the mean of the values right of the transition value should be large enough. Large enough means more than 0.3 times the standard deviation of all the values in the scan. If it is not large enough, we set the transition value to the lowest value of the gate that has been scanned following a similar reasoning as for the previous check. In this scenario we assume that the scan range started at a voltage around 0~mV, and thus that no significant change in the measured current corresponds to a channel that was always open. 
		
	\end{itemize}
\end{itemize}

\subsection{Analysis of single dots}
\label{sec:analysis_of_single_dots}
\begin{figure}[ht]
	\centering
	\begin{tabular}{c c}
		(a) & (b) \\
		\includegraphics[width=0.2\textwidth]{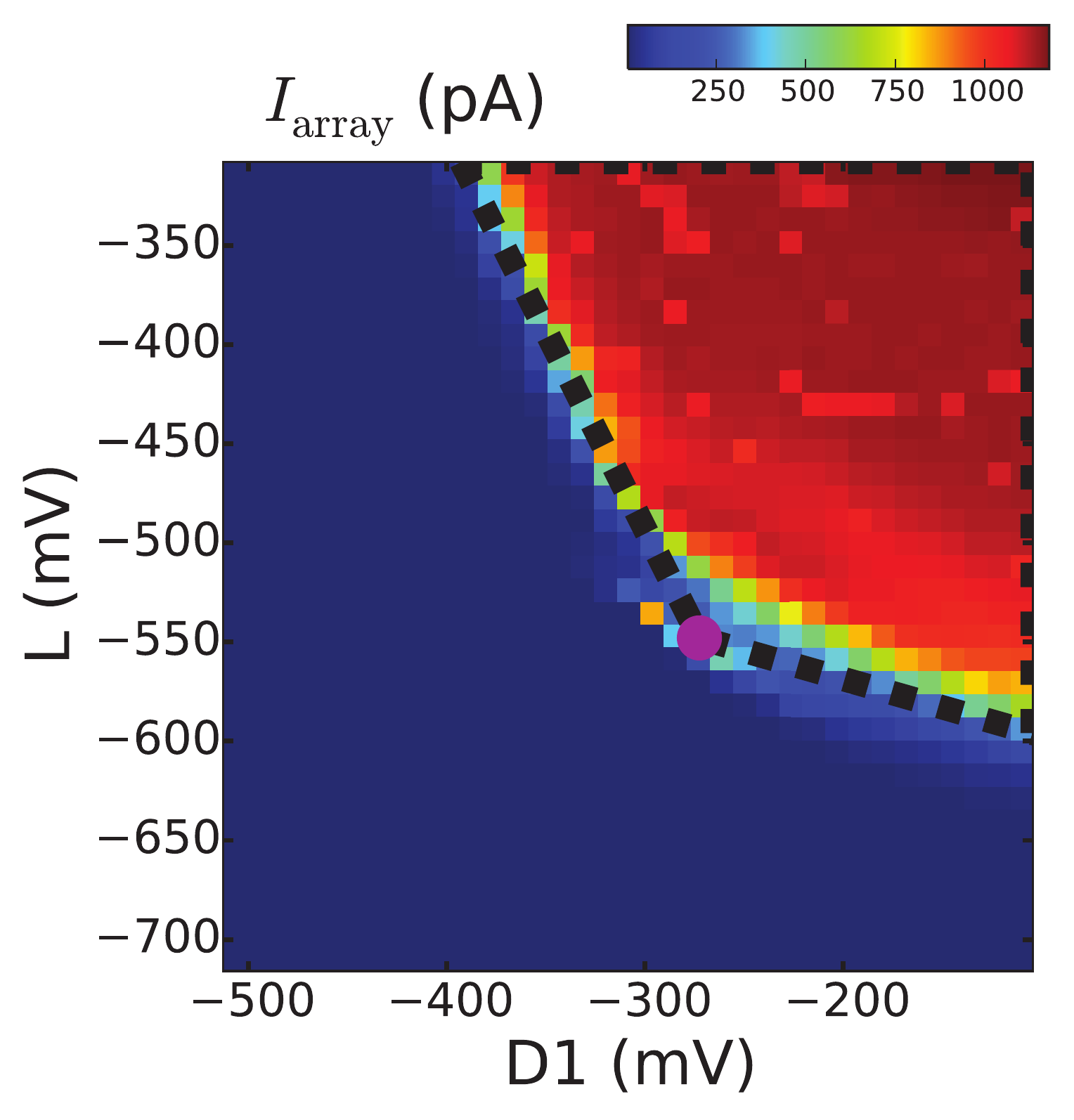} & \includegraphics[width=0.2\textwidth]{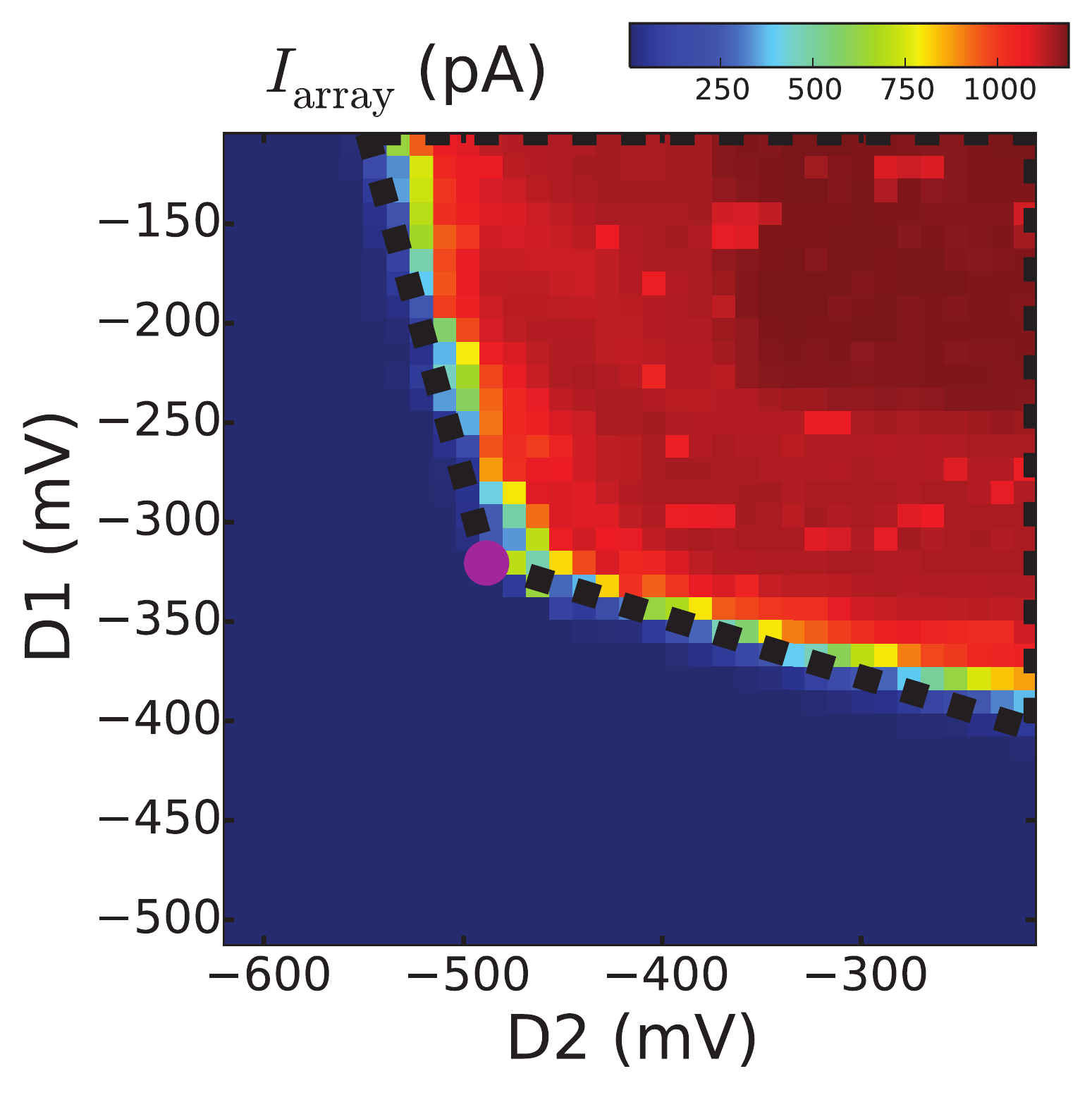} \\ 
		
		(c) & (d) \\
		\includegraphics[width=0.2\textwidth]{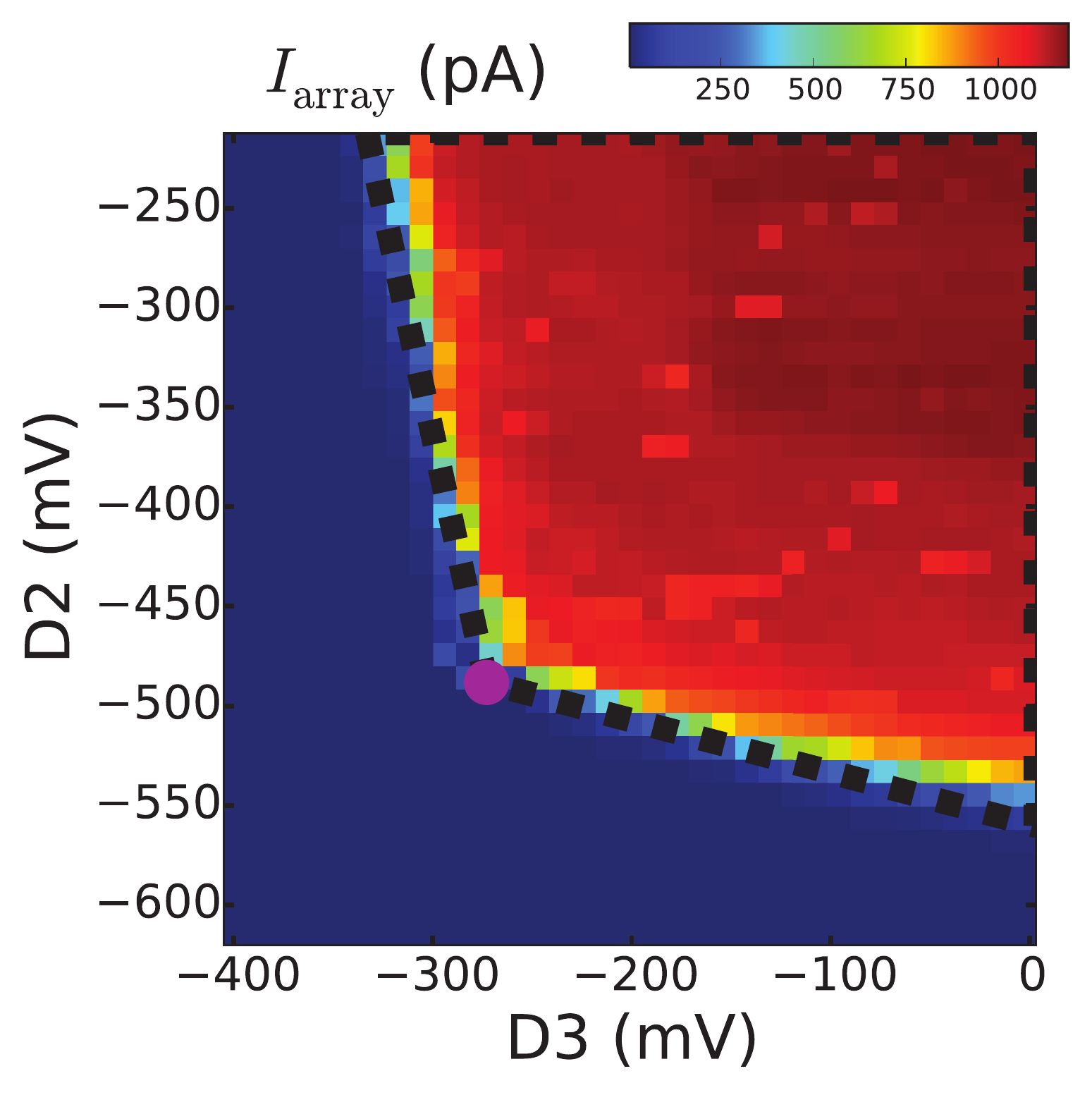} &
		\includegraphics[width=0.2\textwidth]{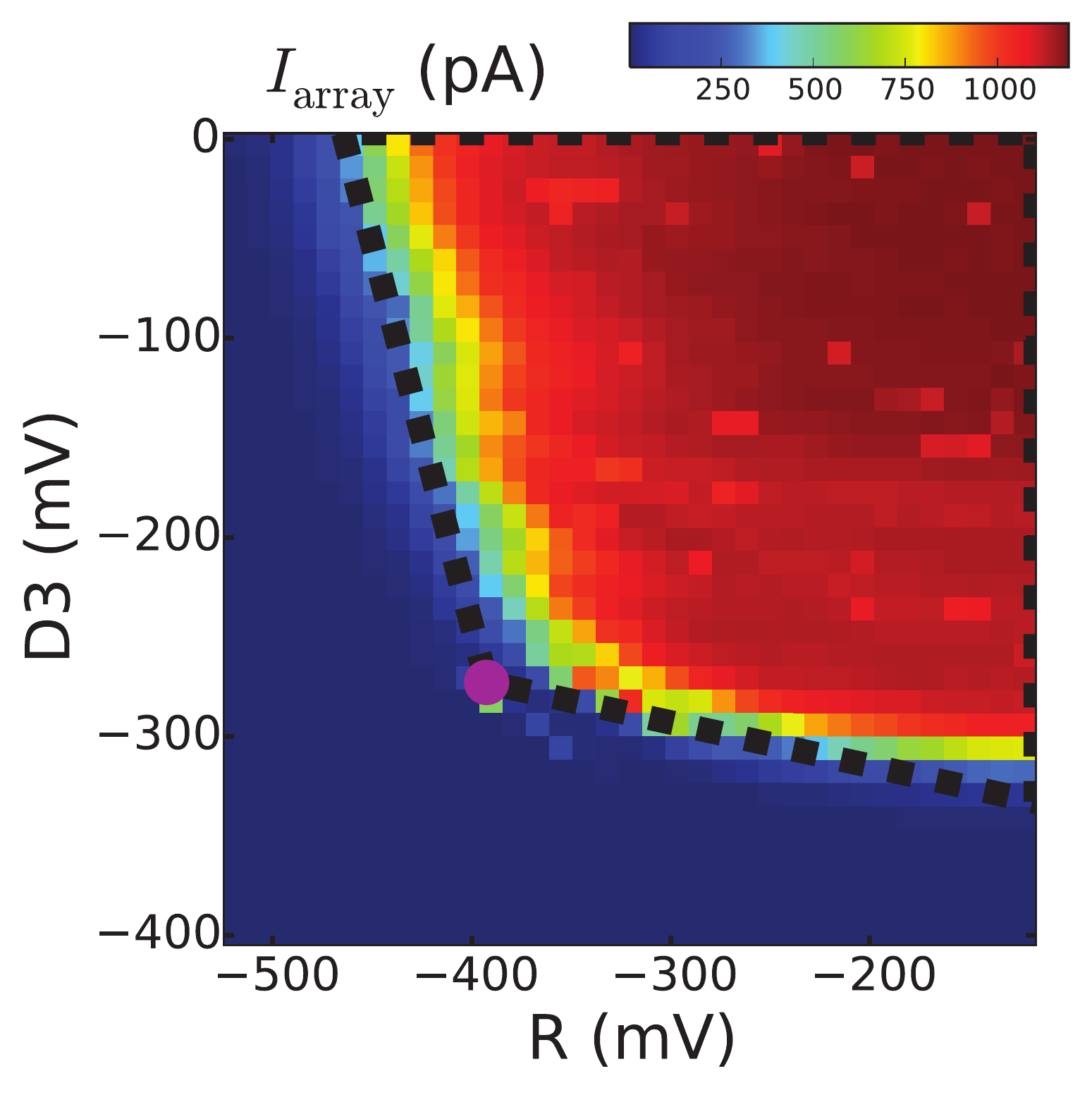} \\ 
	\end{tabular}
	\caption{(a-d) Four single dots are created by sweeping the tunnel barriers for each dot whilst keeping the plunger fixed. The outcome is fitted to a tetragon (dashed black lines) to find the best gate values to search for Coulomb peak formation.}
	\label{fig:quadrilateral_fit}
\end{figure}

\begin{figure}[ht]
	\begin{center}
		\includegraphics[width=.7\columnwidth]{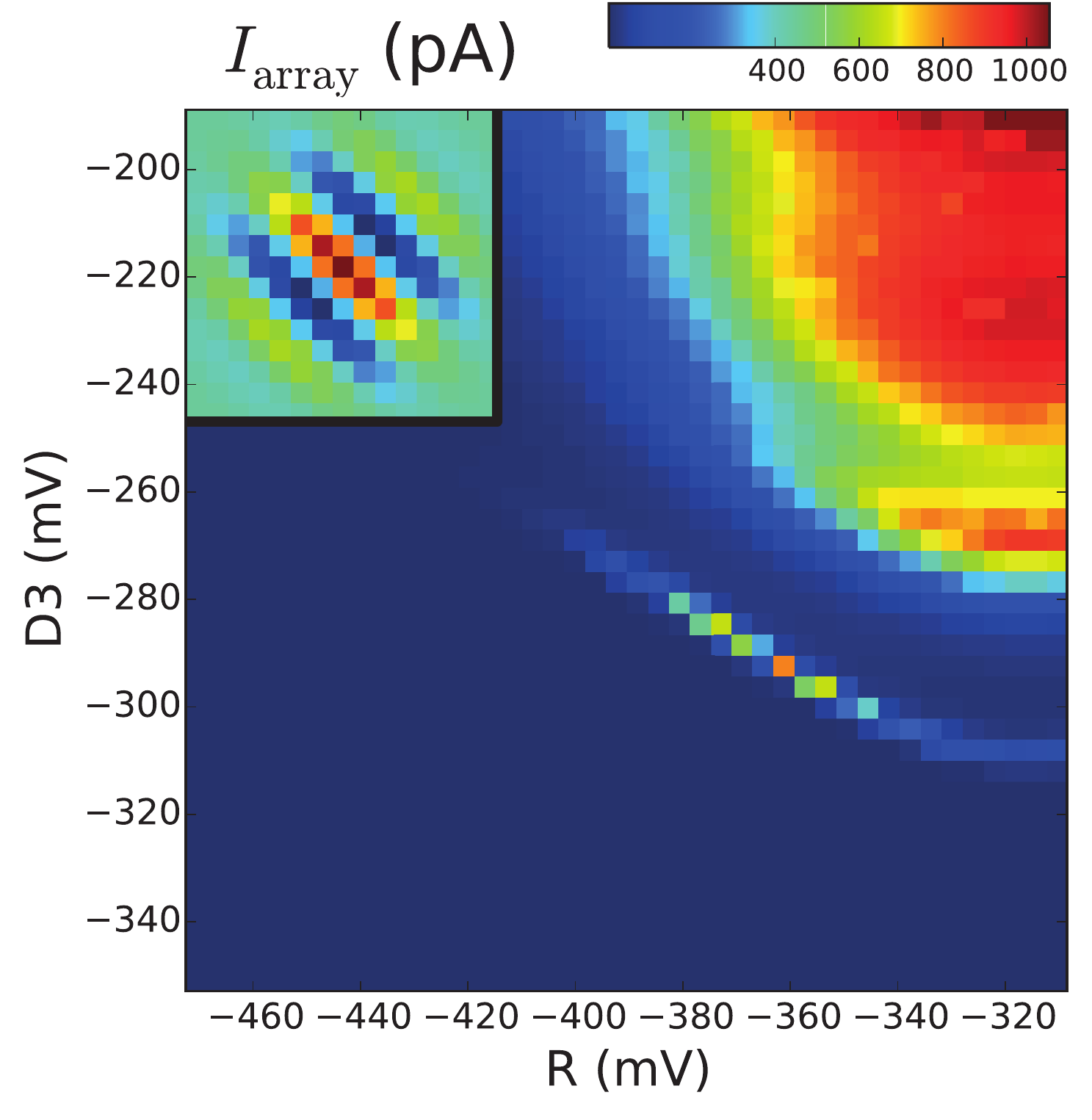}
		\vspace{-.5cm}
	\end{center}
	\caption{Scan of the rightmost single dot where the tunnel barriers are varied. The inset depicts the Gabor filter used to determine the location where Coulomb peaks are formed, see Fig.~\ref{figure:gaborresponse} for the filter response.}
	\label{figure:gaborinset}
\end{figure}

As described in the main text the initial analysis of a 2D scan is performed by fitting a tetragon to the image. The bottom-left corner point of this tetragon gives a good indication of the position of the area where Coulomb peaks are visible. See the magenta points in Figs.~\ref{fig:quadrilateral_fit}(a-d).
The next step is to make a more detailed scan of the area. In this scan we search for the Coulomb peaks using a Gabor filter~\cite{wiki:GaborFilter, Daugman1985}. A Gabor filter is a sinusoidal wave multiplied by a Gaussian function. The sinusoidal wave seeks out features that show contrast with their environment. We define
\begin{align}
G(x,y,\lambda,\theta,\psi,\sigma, \gamma) = \exp\left(- \frac{{x'}^2+\gamma^2{y'}^2}{2\sigma^2} \right) \cos( 2\pi x'/\lambda + \psi)
\label{definition:Gabor}
\end{align}
with $x' = \cos(\theta)x+\sin(\theta)y$, $y' = -\sin(\theta)x+\cos(\theta)y$.
We create a Gabor filter with the following parameters: The orientation is set to
$\theta=\pi/4$, standard deviation of Gaussian $\sigma=12.5$~mV, $\gamma=1$, $\lambda=10$~mV, $\psi=0$. A rectangular image patch of size $40\times40$~mV$^{2}$ is created using this Gabor function (see inset~\figureref{figure:gaborinset}).

\begin{figure}[ht]
	\begin{center}
		\includegraphics[width=.7\columnwidth]{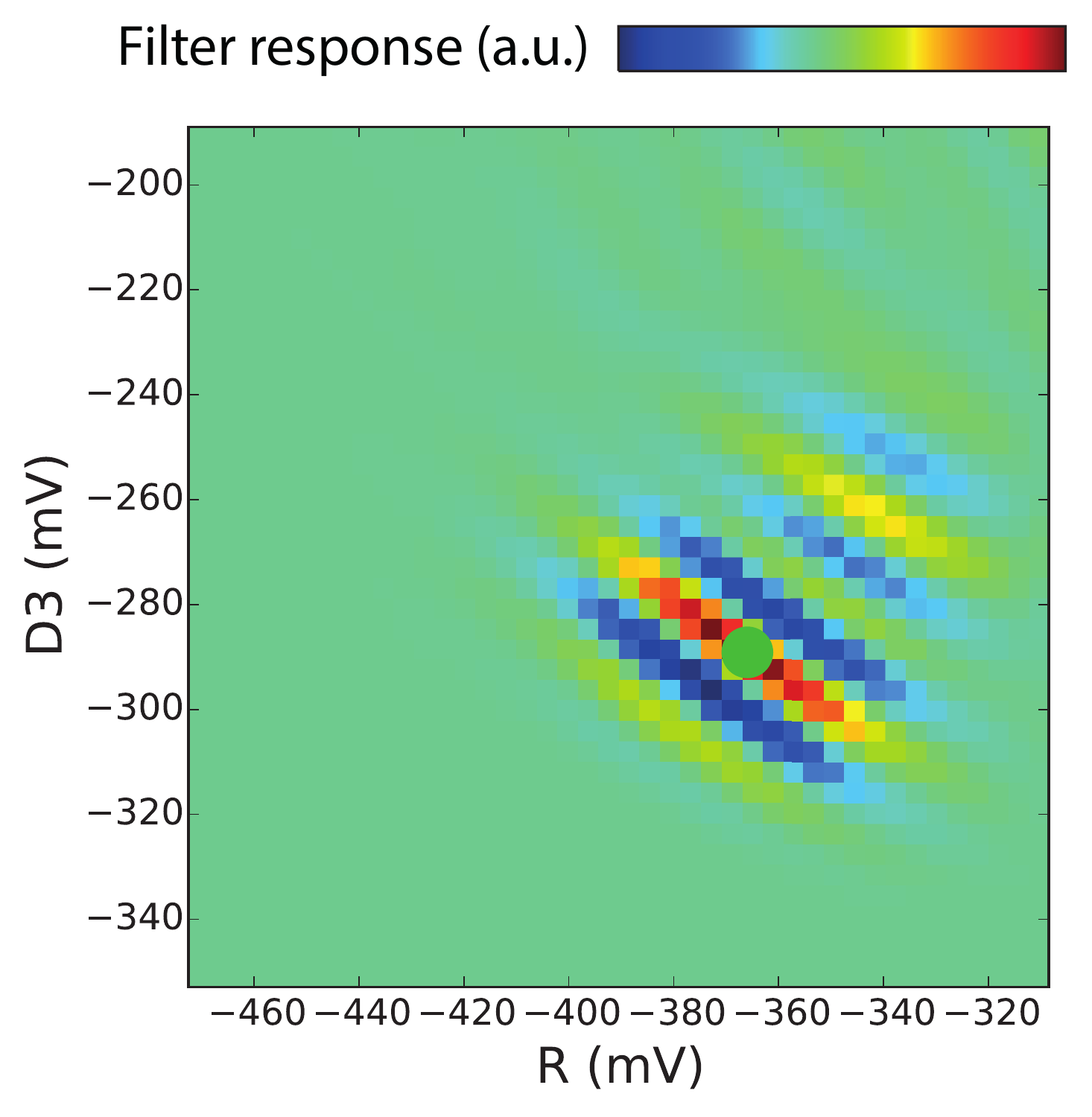}
		\vspace{-0.5cm}
	\end{center}
	\caption{Response of the Gabor filter for the data shown in Fig.~\ref{figure:gaborinset}. Green point is selected as the best Coulomb peak.}
	\label{figure:gaborresponse}
\end{figure}

The response of the Gabor filter to the 2D scan of~\figureref{figure:gaborinset} is shown in~\figureref{figure:gaborresponse}.
\footnote{We use the OpenCV function $\texttt{matchTemplate}$ with method \texttt{TM\_CCORR\_NORMED}.}
From the response figure it is clear that there is a peak (red color) at the location of the Coulomb peak.
To extract the precise location of the peaks we threshold the image with a value automatically determined from the response image and determine the connected components, i.e. the pixels that together constitute a relevant feature. The center of the component furthest into the direction of the closed region is selected as the best Coulomb peak (green point).
The concept of connected components for a binary image is standard in the computer vision community. We find these using OpenCv, see \url{http://docs.opencv.org/3.0.0/d0/d7a/classcv_1_1SimpleBlobDetector.html}, but any implementation will return the same results. If there are more than two Coulomb peaks there are two possibilities:
\begin{enumerate}
	\item Both peaks are converted into two separate connected components. Then, the blob furthest into the direction of the closed region of the scan is selected.
	\item One of the two peaks might be much stronger than the other. In that case the other peak is below the threshold selected by the algorithm and the other peak will be not be visible.
\end{enumerate}
In our experiments we observed that for the single dots at most a couple of Coulomb peaks are visible. Scans of the region of \figureref{figure:gaborinset} with a charge sensor instead of the current through the array confirm that the last Coulomb peak visible in the image is indeed the Coulomb peak corresponding to the zero-to-one electron transition. This behavior is typical for this specific gate design and influences the choice for the plunger gate values for the DQD scans such as shown in Fig.~\ref{fig:Fig3} to find the single-electron regime, see also section \ref{sec:tuning_double_dot}.

\subsection{Selection of Coulomb peaks}
\label{section:coulumbtuning}
\begin{figure}[ht]
	\begin{center}
		\includegraphics[width=.8\columnwidth]{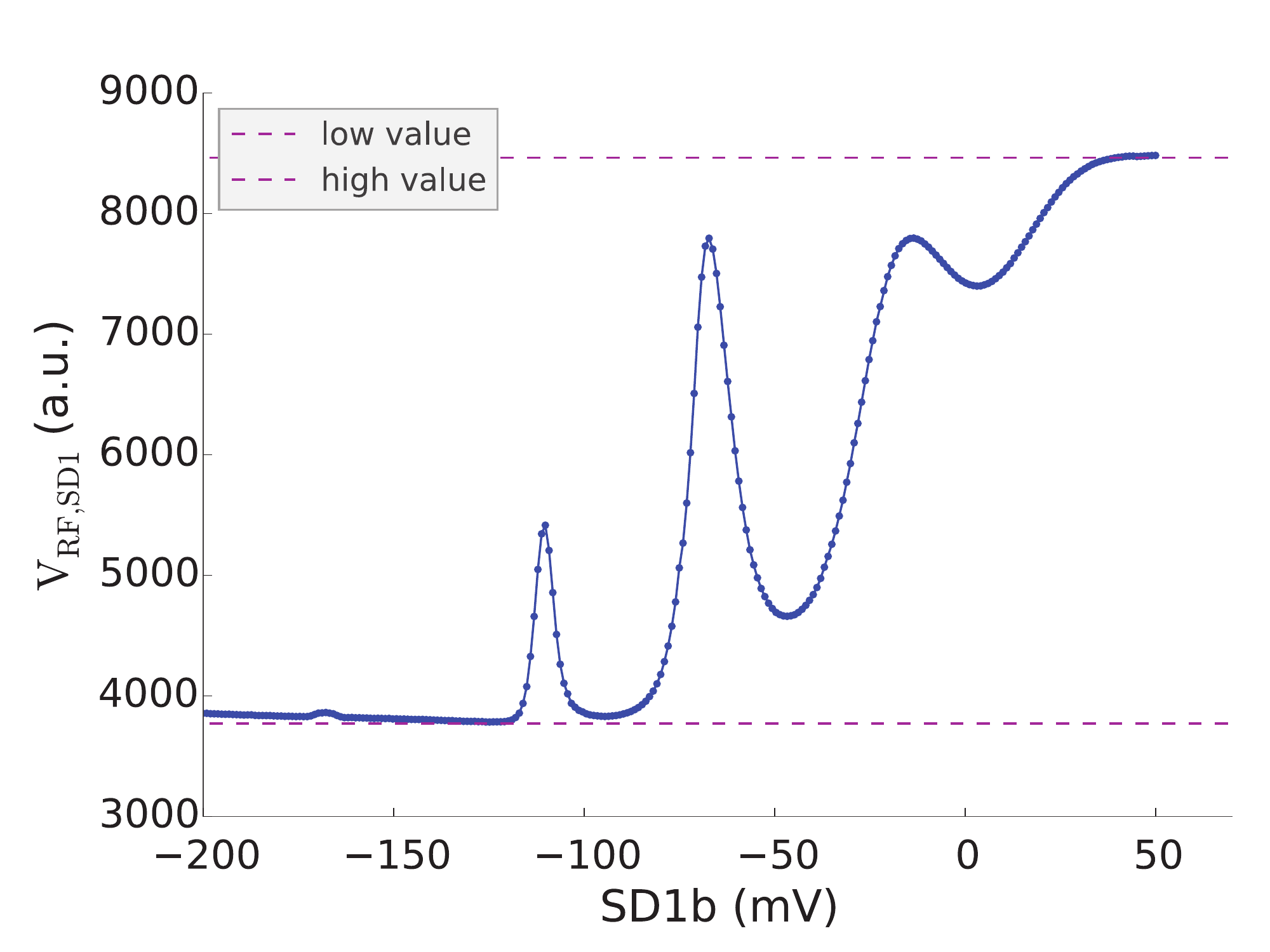}
		\vspace{-.3cm}
	\end{center}
		\caption{Scan of the reflectance $V_{\mathrm{RF,SD1}}$ for varying plunger gate $SD1b$.}
	\label{figure:cexamplescan}
\end{figure}

In this section we describe how the selection of Coulomb peaks for the SD is performed. We start with a scan of the plunger gate $SDxb$ ($x$ is 1 or 2) in a configuration for which Coulomb peaks can be expected. In the scan we determine the lowest and highest values (meaning the channel is completely closed or completely open) using a robust minimum and maximum function (see section~\ref{sec:pinch_off} for details). These values are indicated as dotted horizontal lines in~\figureref{figure:cexamplescan}.

\begin{figure}[ht]
	\begin{center}
		\includegraphics[width=.8\columnwidth]{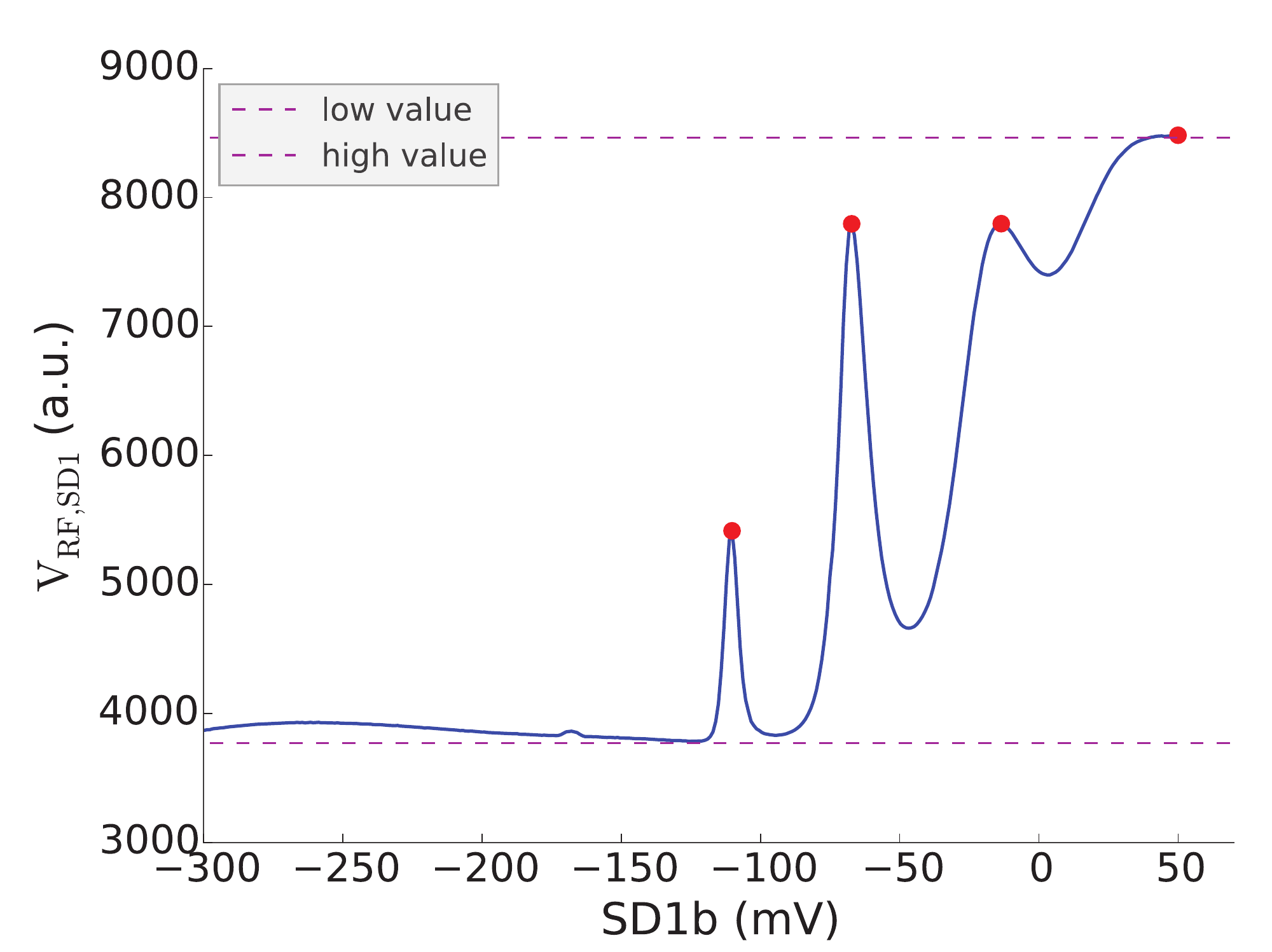}
				\vspace{-.3cm}
	\end{center}
		\caption{Detected peaks in the data of Fig.~\ref{figure:cexamplescan}.}
	\label{figure:cexamplepeaks}
\end{figure}

\begin{figure}[ht]
	\begin{center}
		\includegraphics[width=.8\columnwidth]{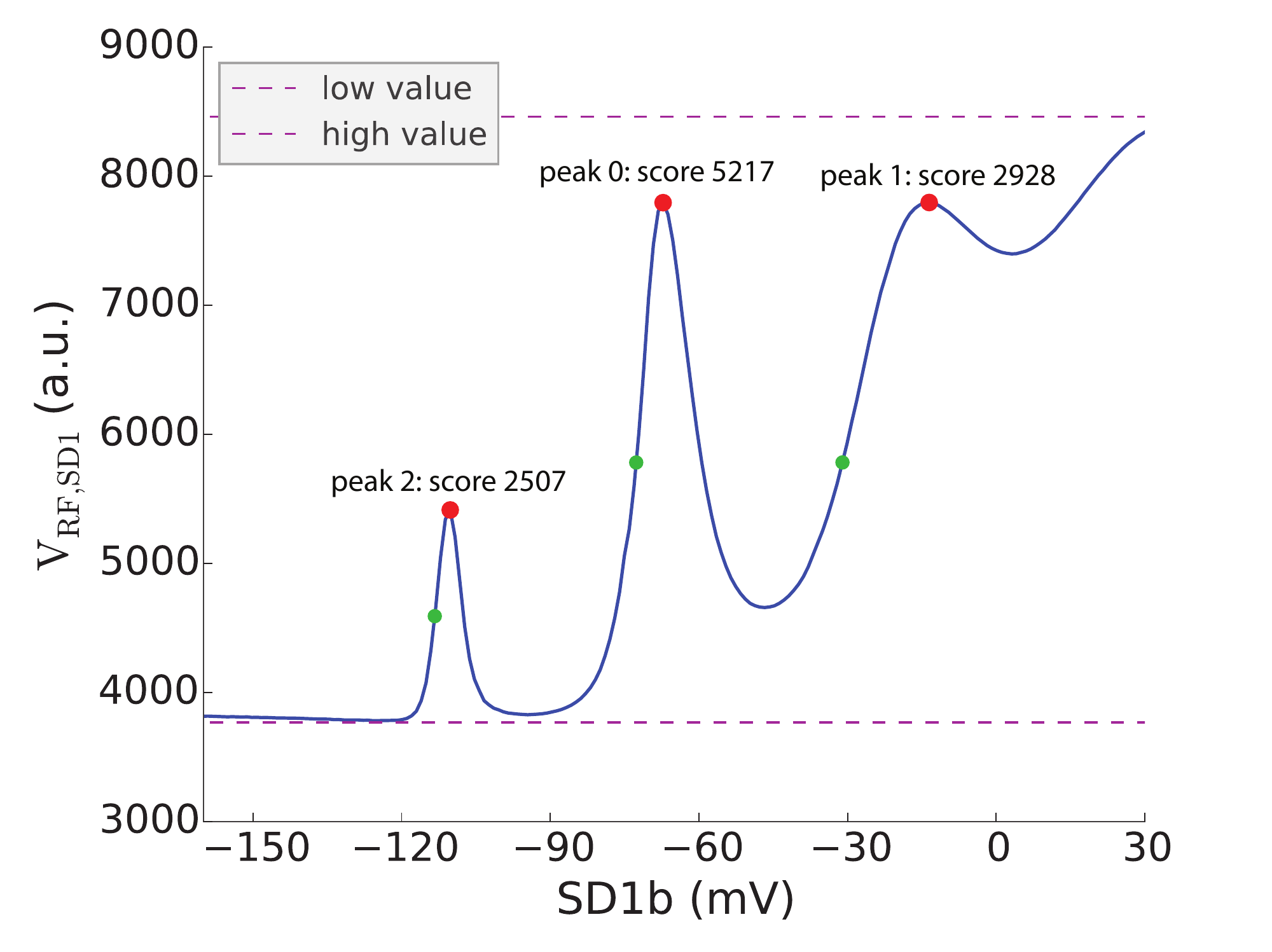}
		\caption{Final result of indexing the data from Fig.~\ref{figure:cexamplescan}. The peaks are plotted in red, the point at the peak half-height on the left is plotted as a green dot.}
		\label{figure:cexamplefinal}
	\end{center}
	
\end{figure}

The peaks in the data are detected by selecting the local maxima in the plot. All peaks below a minimum threshold are discarded.\footnote{We use \texttt{scipy.ndimage.filters.maximum\_filter1d} with a size parameter of 12~mV.}
For each of the peaks the position of the peak half-height on the left and right side is determined. Also the bottom of the peak is determined (see details at the end of this section). From these values we can determine the peak properties such as the height and the half-width of the peak. Finally the peaks are filtered based on the peak height and overlapping peaks are removed (see details at the end of this section) leading to the detected peaks shown in Fig.~\ref{figure:cexamplepeaks}.\\
After this filtering step the peaks are ordered according to a score. For various applications we can define different scores. In this work the SD-peaks are primarily selected for proper charge sensing. For a good charge sensitivity we need a large peak with a steep slope. We then tune the SD to the position at half-height on the left of the highest-scoring peak. The scoring function we used is
\begin{align}
\textrm{score} &= \textrm{height} \frac{2}{1+\textrm{hw}/\textrm{hw}_0}  
\end{align}
The value of $\textrm{hw}_0$ is a scaling parameter determining the typical half width of a Coulomb peak. In our experiments we used $\textrm{hw}_0=10$~mV. In our experience, this scoring represents a reasonable trade-off between the height and the slope of a peak. The result is shown in Fig.~\ref{figure:cexamplefinal}.

\indent \textit{Details to detect the bottom left of a peak}: the $x$-coordinate of a peak, $x_{\mathrm{peak}}$, has already been determined by selecting the local maxima in the data. To find the $x$-coordinate of the bottom on the left side of the peak, $l$,  the following steps are performed after smoothing the data:
\begin{enumerate}
	\item Search for the $x$-coordinate, $x_{\mathrm{bottom~low}}$, of the minimum value (bottom low) in the range ${[ x_{\mathrm{peak}}-3 \times thw, x_{\mathrm{peak}}]}$. The variable $thw$ is a measure for the typical half width of a peak and is set to $\sim10$~mV.
	\item  Starting from $x_{\mathrm{bottom~low}}$, scan from left to right and select the first datapoint that fulfills the following two conditions: (1) the slope is positive, and (2) the $y$-value is larger than `bottom low + 10\% of the peak height'.
\end{enumerate}
This method does not require a specific fitting model, and also works well for asymmetric Coulomb peaks.\\

\indent \textit{Details of the filter to remove overlapping peaks}: for each peak we have the position of the bottom on the left ($l$) and the top of the peak ($p$). For two peaks the overlap is defined using the intersection of the intervals $[l_1, p_1]$ and $[l_2, p_2]$. The length of an interval $L$ is denoted as $||L||$. The overlap ratio is then equal to ${ || \pi([l_1, p_1], [l_2, p_2]) || } / {\sqrt{ || [l_1, p_1] || || [l_2, p_2] ||}}$. To make the overlap a bit more robust we use a smoothed version of this formula using Laplace smoothing:
\begin{align*}
s &= \frac{ 1+ || \pi([l_1, p_1], [l_2, p_2]) || }{1+\sqrt{ || [l_1, p_1] || || [l_2, p_2] ||} } .
\end{align*}
When the overlap $s$ between two peaks is larger than a threshold (0.6), then the peak with the lowest score is removed.

\subsection{Tuning and analysis of a double dot}
\label{sec:tuning_double_dot}
The main text describes how we set the gate values for the tunnel barriers of each double dot using the information of the single dot scans. For the plunger gates of the double dot an extra compensation factor is added. When each single dot is formed, the dot-barrier gate of its neighbor is kept at zero Volt. When next making a double dot, these dot-barrier gates are activated and shift the electrochemical potential of their neighbor, for which we compensate with the corresponding plunger voltage. This compensation factor is determined heuristically. For a double dot with gates $L-P1-M-P2-R$ the compensation values for $P1, P2$ are
\begin{align*}
\left( -\phi R, -\phi L  \right) ,
\end{align*}
with $\phi=0.1$. See Table~\ref{tab:example_double_dot_settings} for an example. In future experiments we plan to use the capacitive-coupling information from the single dot scans in order to create more precise compensation values for the tunnel barrier gates. The exact values of the plunger gates are not very important, since we will make a scan of the double dot using the plunger gates. A good initial guess does reduce the measurement time.
\begin{table}
\setlength{\tabcolsep}{3.5pt}
\begin{tabular}{r|r|r|r}
\bfseries Gate & \bfseries Left dot & \bfseries Right dot & \bfseries Double-dot  \\
\hline
$L$ & -539.9 &  & -539.9  \\
$P1$ & -80.0 &  & -33.0  \\
$D1$ & -285.9 & -327.3 & -306.6  \\
$P2$ &  & -80.0 & -26.0  \\
$D2$ &  & -469.7 & -469.7  
\end{tabular}

\caption{Example for double dot settings. All values are in mV.}
\label{tab:example_double_dot_settings}
\end{table}

The important structures in the scan of a double dot are the charging lines and the crossings of two charging lines. To determine the locations of the crossings in the image we create a template for such a crossing. We then search for crossings using template matching. The response to the template is thresholded and local maxima are calculated. The template of the crossing consists of 4 lines at angles $\pi/8$, $3\pi/8$,  $9\pi/8$ and $11\pi/8$ (radians) that are separated by a distance of 1.5~mV at 45 degrees that represents the interdot capacitive coupling (see inset of Fig.~\ref{fig:Fig3} of the main text). The width of these lines ensures that experimentally measured crossings still overlap with the template despite unavoidable small variations in the interdot capacitive coupling and the lever arms between gate voltage and electrochemical potential, which affect the slope of the transitions.

The final step consists of checking whether extra charging lines are visible in a region of $\sim$70$\times$70~mV$^{2}$ to the side of more negative gate voltages. The size of this region should be larger than the charging energy of each dot in mV. The topright corner of the 70$\times$70~mV$^{2}$ area is located -10~mV southwest of the most bottom-left cross. We slightly extend this region on top and on the right to reduce the probability that a charging line is missed. If the total region falls outside the scan range of the data, the algorithm reduces the size of this region accordingly (alternatively, one could take data over a larger gate voltage range). The algorithm could then draw the wrong conclusion. When the region that results from clipping at the border of the scan range is smaller than 40$\times$40~mV$^{2}$, the algorithm will stop and output that it cannot properly determine whether the single-electron regime has been attained. In the typical case that the region is large enough, we first smoothen the data within this region. We subtract the smoothed data from the original data and check whether the resulting pixel values fall above a certain threshold that is proportional to the standard deviation of the smoothed dataset. If at most one pixel value is larger than the threshold, the algorithm classifies the dataset as `single-electron regime'.

\end{document}